\documentclass[12pt]{article}
\usepackage[square]{natbib}
\usepackage{graphicx}
\usepackage{setspace}
\usepackage{authblk}

\title{Paleo-heat flux estimates across Enceladus derived from flexural response of the ice shell}
\author[a]{Chelsea Eakins}
\author[a]{Francis Nimmo}
\author[b]{Paul Schenk}
\affil[a]{University of California, Santa Cruz}
\affil[b]{Lunar and Planetary Institute}
\date{}

\usepackage{natbib}
\usepackage{amsmath}
\usepackage{amssymb}
\usepackage{appendix}
\usepackage{multirow}
\usepackage{float}
\usepackage{pdflscape}

\usepackage{rotating}

\usepackage[bf]{caption}
\usepackage{color}
\usepackage{multirow}

\renewcommand{\title}[1]{\bc {\Large \bf #1} \ec}

\newcommand{\bc}{\begin{center}}
\newcommand{\ec}{\end{center}}
\newcommand{\be}{\begin{equation}}
\newcommand{\ee}{\end{equation}}

\newcommand{\beginsupplement}{%
        \setcounter{table}{0}
        \renewcommand{\thetable}{S\arabic{table}}%
        \setcounter{figure}{0}
        \renewcommand{\thefigure}{S\arabic{figure}}%
     }

\usepackage{csvsimple} 
\usepackage{booktabs} 
\usepackage{siunitx} 
\usepackage{pgfplotstable} 
\usepackage[T1]{fontenc} 
\usepackage{pgfplots}
\pgfplotsset{compat=1.18}

\begin{document}

\maketitle

Saturn's moon Enceladus has active plumes suggestive of locally high heat fluxes at present, and the present-day heating is reasonably well characterized. However, the moon's thermal history is not well understood. 
In this study, we model flexural deformation of chasmata and dorsae in various regions of Enceladus to find the effective elastic thickness of the ice shell and constrain temporal and spatial variations in heat flux.
We find paleo-heat fluxes of $60-\geq350~\rm mWm^{-2}$, with most clustered around $125-200~\rm mWm^{-2}$.
Present day measured heat fluxes at the South Polar Terrain (SPT) are comparable to our estimated paleo-heat fluxes in non-SPT regions. Our paleo-heat fluxes, however, are significantly higher than the corresponding present day heat fluxes for the same locations inferred from shell thickness maps.
This suggests that regions outside the SPT experienced a past episode of heating (heat fluxes in excess of $100~\rm mWm^{-2}$) followed by a cooling and thickening ice shell.
A relatively constant paleo-heat flux for terrains of different ages (determined from crater densities) supports the "resonance locking" hypothesis of dissipation in Saturn and might imply a long-lived resonance between Enceladus and Dione.
Our results suggest that Enceladus's tidal heating appears to have been consistent over time. Consistent peak heating at Enceladus could support a long-lived subsurface ocean, which has important consequences for astrobiology and the habitability of the outer solar system.

\section{Introduction}
\label{intro}

Enceladus is a small icy moon of Saturn, with active plumes and excess heat detected at its south pole \citep{porco_cassini_2006,spencer_cassini_2006,hansen_plumes_2006,dougherty_magnetometer_plumes_2006}. It has a diameter of approximately $500~\rm km$, an ice shell average thickness ranging from $19-24~\rm km$ \citep{hemingway_mittal_2019}, and a subsurface ocean \citep{thomas_shape_2016}. As a consequence of the ocean and the present-day geological activity, Enceladus is considered a body of high astrobiological interest \citep{mckay_biomarkers_2008,mckay_icymoons_2018}.

Understanding Enceladus's long-term heat budget is crucial to understanding its potential habitability \citep{cockell_comp_habitability_2024}. But although its present-day heat budget is reasonably well characterized \citep{howett_2011,hemingway_mittal_2019,miles_npheat_2025}, quantities like the ocean lifetime and duration of tidal heating are very uncertain. As summarized in \cite{nimmo_dissipation_2023}, both these quantities depend on the orbital evolution of Enceladus and the manner in which Saturn transfers energy and angular momentum to the satellites.

One way of determining Enceladus's long term thermal evolution is to look for evidence of ancient deformation, which can record an ancient heat flux (e.g. \citep{giese_2008,bland_2007}). By comparing the spatial and temporal variation in heat flux, it is then possible to place constraints on thermal evolution. This is the approach we take in this paper.

\subsection{Geology}
\label{intro geol}

Enceladus is characterized by four primary terrains: the leading hemisphere terrain (LHT), trailing hemisphere terrain (THT), cratered plains, and the South Polar Terrain (SPT) \citep{crow-willard_mapping_2015,patterson_icymoons_2018}. The LHT and THT are tectonized regions centered about the leading and trailing points respectively, with chasmata and dorsae the most prominent features. 
The cratered plains cover nearly $40\%$ of the surface, characterized by modest tectonization and higher crater densities than other terrains. 
The cratered plains encircle the LHT and THT, which are bounded by curvilinear belts that feature a complex combination of extension, contraction, and shearing. One potential formation mechanism for the LHT and THT is a period of upwelling and ice shell thinning followed by downwelling and ice shell thickening \citep{crow-willard_mapping_2015}.

The SPT is the youngest terrain on Enceladus, with active plumes detected by \textit{Cassini} \citep{porco_cassini_2006,hansen_plumes_2006,dougherty_magnetometer_plumes_2006} associated with the tiger stripes, four semi-parallel fractures potentially extending down to the subsurface ocean. 
This region is the locus of geologic activity on Enceladus, and it has been suggested that plume material resurfaces terrain outside the SPT \citep{kirchoff_crater_2009,kinczyk_cratered_terrains_2024}.
Encircling the outside of the SPT is the southern curvilinear region, also called the SPT margins \citep{yin_spt_2015,crow-willard_mapping_2015}. 
The SPT margins are similar to the curvilinear regions bounding the LHT and THT, with extensional and compressional features. 
\cite{yin_spt_2015} suggest that the SPT and the margins formed from gravitational spreading, to explain compressional features near the THT and extensional features near the LHT (on opposite sides of the SPT).

\subsection{Context and Outline}
\label{intro outline}

Previous studies have estimated both modern and ancient heat fluxes on Enceladus, finding present-day values of $50-500~\rm mWm^{-2}$ \citep{hemingway_mittal_2019,howett_2011,spencer_cassini_2006,miles_npheat_2025} and ancient heat fluxes ranging from $100-300~\rm mWm^{-2}$ \citep{bland_2007,bland_crater_relaxation_2012,giese_2008,leonard_2021}. 
Present-day conductive heat fluxes have been estimated from shell thickness maps \citep{hemingway_mittal_2019}. \textit{Cassini} infrared measurements taken of the South Polar Terrain and North Pole constrain the present-day heat loss at Enceladus \citep{howett_2011,spencer_cassini_2006,miles_npheat_2025}. 
Past heating has been estimated through both crater relaxation and though analyzing flexural response of the ice shell. In the northern latitudes of the trailing hemisphere, viscous relaxation of craters has been used to estimate ancient heat fluxes \citep{bland_crater_relaxation_2012}. Multiple studies have estimated elastic thicknesses and inferred heat fluxes in the trailing hemisphere terrain using flexural deformation \citep{giese_2008} and spacing of extensional features \citep{bland_2007}. In the leading hemisphere, one study inferred lithospheric thickness from stresses of rotating blocks \citep{leonard_2021}.
More detail is provided on the comparison between our findings and those of previous studies in Sections \ref{present hfs},\ref{past hfs}.

We take an approach of modeling the flexural response of the ice shell on Enceladus (discussed further in Section 
\ref{methods}) consistent with approaches taken on other icy satellites such as Europa, Ganymede, Tethys, Dione, etc. \citep{billings_europa_2005,hurford_europa_flexure_2005,nimmo_ganymede_2002,beddingfield_tethys_2023,giese_tethys_2007,hammond_dione_flexure_2013,beddingfield_ariel_2022,beddingfield_titania_2023,conrad_2019}.
In this study, we attempt to map global variations in paleo-heat fluxes using flexural modeling of the ice shell deflection. 
 
We find paleo-heat fluxes that are comparable to the present-day heat fluxes experienced at the South Polar Terrain. In combination with a lack of correlation between local terrain age and heat flux magnitude, this suggests that the peak heat flux on Enceladus has not changed significantly over time. 
Our results are supportive of the resonance locking model of dissipation in Saturn \citep{fuller_resonance_2016}, which if true, would suggest a long-lived mean motion resonance between Dione and Enceladus along with consistent peak heat fluxes on Enceladus over time \citep{nimmo_icymoons_2018}. 
This heating could sustain a long-lived subsurface ocean and has implications for the habitability of Enceladus \citep{cockell_comp_habitability_2024}.

The rest of this paper is organized as follows. In section 2, we outline the methods, describing the flexural models we use and the associated uncertainty, and relating our best fit elastic thicknesses to paleo-heat fluxes.
Section 3 provides the results of our flexural models: the best fit elastic thicknesses across Enceladus and the corresponding paleo-heat flux estimates.
Then in section 4, we compare our estimated paleo-heat fluxes to present-day measured values and both present-day and paleo-heat flux estimates. In section 4, we also consider the relationship between our estimated paleo-heat fluxes and terrain age.
Finally, section 5 discusses the implications of our results on the heating history of Enceladus, with consequences for both the orbital evolution and the long-term stability of a subsurface ocean.

\section{Methods}
\label{methods}

\begin{figure}
    \centering
    \includegraphics[width=0.9\textwidth]{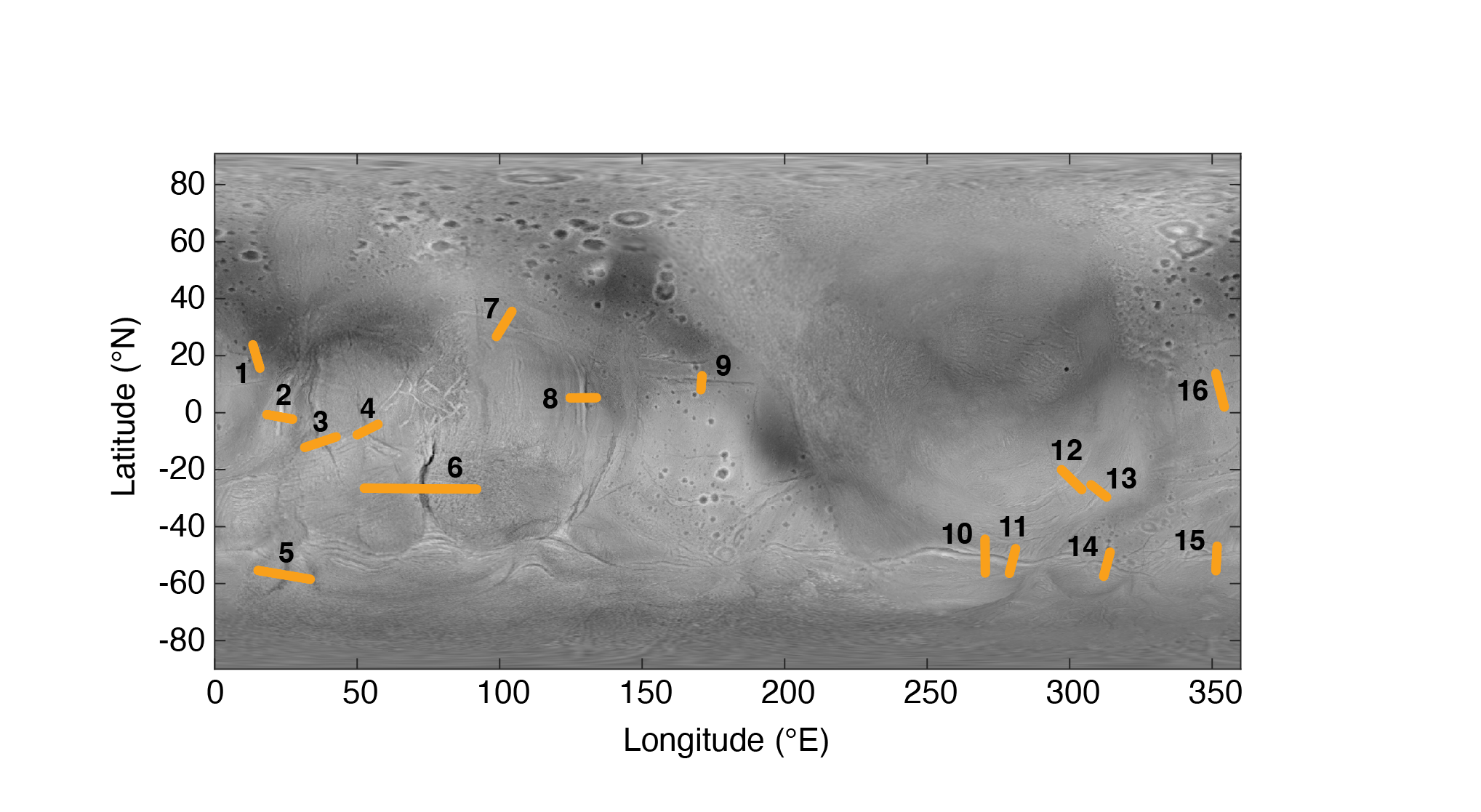}
    \caption{Digital elevation map (DEM) of Enceladus from \citep{schenk_2024} relative to a triaxial ellipsoid. Elevation ranges from +2.3391 km (white) to -2.7330 km (black). The orange lines represent averages of elevation profiles over chasmata and dorsae across the surface. Each feature is indicated with a number, as shown in Table \ref{tab:feature tab}.}
    \label{fig:FeatureMap}
\end{figure}

We use a global digital elevation model (DEM) of Enceladus, as seen in Figure \ref{fig:FeatureMap}, generated primarily from stereo analysis \citep{schenk_2024}.
The DEM is constructed from photogrammetric control network (PCN) control point radius estimates and stereogrammetry. The stereogrammetric DEM is constructed from \textit{Cassini} stereo image pairs and mosaics. The PCN is based on a global shape model of Enceladus, a triaxial ellipsoid as specified by \citep{thomas_shape_2016}, and the trajectory of the spacecraft. The DEM data are spaced with a resolution of $200~\rm m$ with vertical uncertainty in elevation ranging from $50~\rm m$ to $250~\rm m$.

As described below (Section \ref{numerical},\ref{analytical}), we follow previous studies \citep{giese_2008,barnett_flexure_2002,conrad_2019} and compare theoretical flexural models with the observed topography to determine the best fit value for the effective elastic thickness of the ice shell.

Using the DEM, we generate multiple $(12-24)$ topographic profiles across features of interest. We average these profiles to minimize short wavelength geologic “noise”, such as fracturing of the surface and cratering. The features were chosen in an attempt to provide global coverage of flexural features across the surface of Enceladus. We selected chasmata and dorsae in the leading and trailing hemispheres, as well as the SPT margins. This selection is limited by the heavily cratered northern hemisphere, which does not have readily detectable flexural features, and by the degree of topographic noise surrounding the features.

To model the topography, we use two methods: a numerical approach where the pre-deflection topography (the load topography $L$) is specified and the effective elastic thickness is varied to best match the topography, 
and an analytical approach, where we construct a universal flexural profile (UFP) \citep{turcotte_geodynamics_2002} to match the observed topography without having to specify the load.
The analytical method is based on flexure at oceanic trenches, assuming a broken plate. The numerical method is based on periodic loading of an unbroken elastic plate, such as the flexure from the Hawaiian volcanoes. Both methods use a Cartesian approximation, excluding membrane stresses. The exclusion of membrane stresses is based on the extensive fracturing observed at Enceladus; such fractures suppress the transmission of these stresses. However, if membrane stresses were important, this would reduce our elastic thickness estimates and result in higher local heat fluxes.

We use the numerical method when we can reasonably assume a pre-deflection load topography. Usually, this means dorsae and chasmata that show similar flexural responses on either side of the feature. We use the analytical method for chasmata where we see two different characteristic flexural responses on either side of the feature, or where we cannot assume a load topography.

We use the analytical method for the South Polar Terrain margins to get two estimates of elastic thickness on either side of the feature. As the margins bound the edges of the South Polar Terrain, we split the chasma into North and South sides (further and closer to the SPT, respectively). The independent flexural modeling on the North and South sides allows for the investigation of potential differences in the ice shell on either side of the margin.

\subsection{Numerical Method}
\label{numerical}

For the numerical method, the chasmata or dorsae are approximated using a triangular load. The width and height of the pre-deflection (load) topography, $L$, are visually specified to best match the observed topography.

We take the Fourier transform of the load topography and calculate the lithospheric deflection, $W$, using the following equation \citep{barnett_flexure_2002}:

\be
W(k) = \frac{-L(k)}{\frac{\rho_m}{\rho_c} + \frac{(\alpha k)^4}{4}}
\ee

In this equation, $L(k)$ is the Fourier transform of the load topography, $\rho_{m}$ is the density of the layer beneath the ice shell, $\rho_{c}$ is the density of the elastic layer and the load, $k$ is the wavenumber, and $\alpha$ is the flexural parameter, given by the following equation \citep{turcotte_geodynamics_2002}:

\be
\alpha = \left(\frac{E\ {T_e}^3}{3\ (1 - \nu^2)\ \Delta\rho \ g}\right)^{1/4}
\ee

The flexural parameter is the characteristic wavelength of the lithospheric response to the load. Here, $T_e$ is the elastic thickness of the lithosphere, $E$ is Young’s modulus for ice, $\nu$ is Poisson’s ratio, and $g$ is the gravitational acceleration on the surface. For these calculations, we used: $E = 9\ \rm GPa$, $g = 0.11\ \rm m/s^2$, $\nu = 0.25$, $\Delta\rho = 1000\ \rm kg/m^3$. The density contrast, $\Delta\rho$, represents the difference in density between the underlying and overlying material, which we take to be the density of water and vacuum, respectively. 

\begin{figure}
    \centering
    \includegraphics[width=1.0\textwidth]{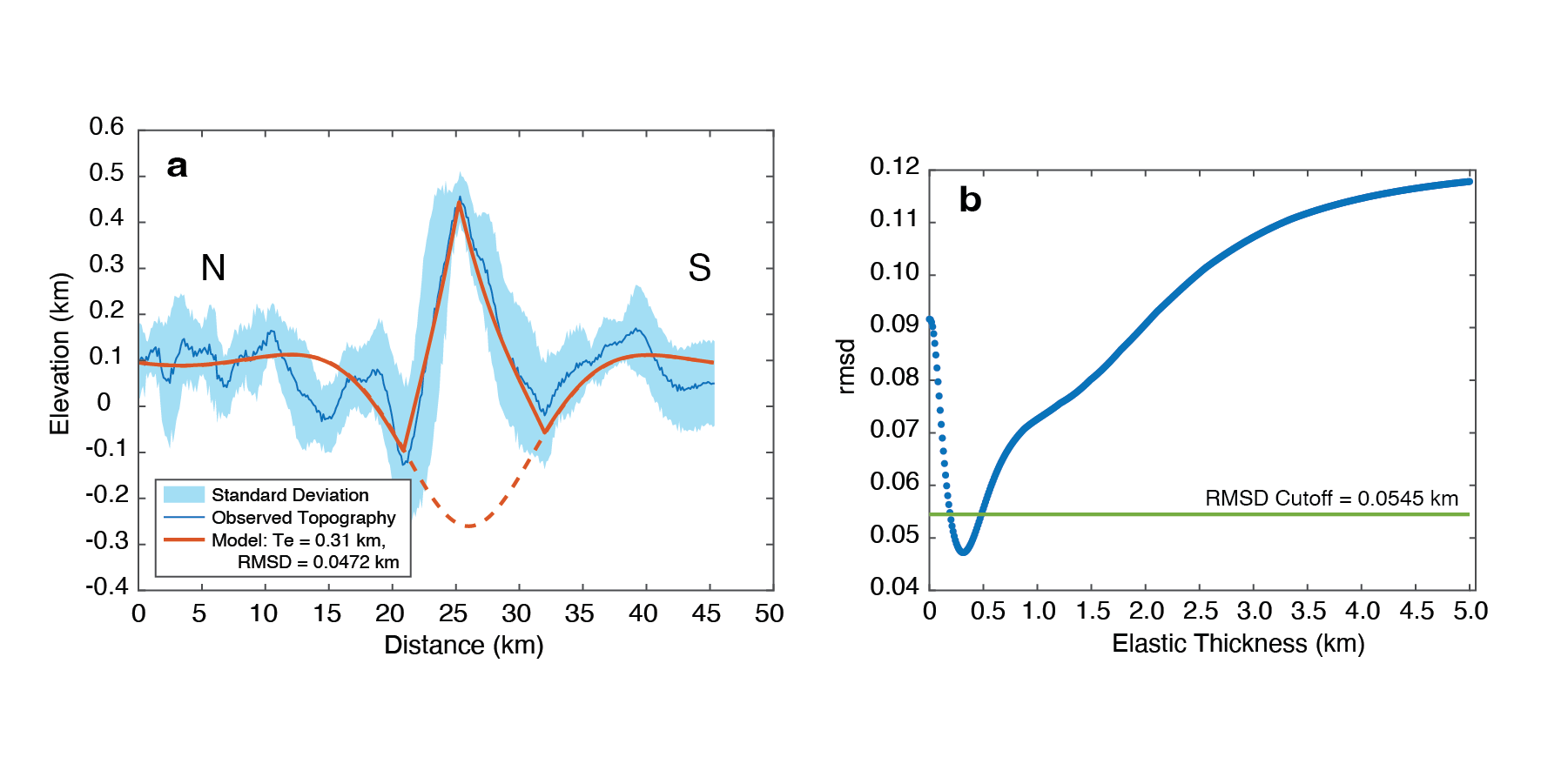}
    \caption{a) Best fit flexural model (orange) plotted with the averaged topography (blue) for feature $\#7$ (Figure \ref{fig:FeatureMap}, Table \ref{tab:feature tab}). The blue shaded region represents the standard deviation of the topographic profiles. The dashed orange line shows the deflection of the given load. This model is constructed using the numerical method, assuming a triangular load with a height of $0.7~\rm km$ and a width of $11.1~\rm km$. The best fit model uses an elastic thickness of 0.31 km. b) Misfit of our model as a function of elastic thicknesses. The minimum misfit has an RMS deviation of 0.047 km. The $95\%$ RMSD cutoff value for the best fit model is 0.0545 km, giving an estimated $T_e$ range of 0.20 to 0.48 km.}
    \label{fig:EXFeature}
\end{figure}

We compute the final topography by summing the deflection and the load topography using the equation $H(k) = L(k)+W(k)$ where $H$ is the final topography, $W$ is the deflection, and $L$ is the load topography, all in the wavenumber domain. Taking the inverse Fourier transform gives the final topography over the spatial domain. We vary the elastic thickness, $T_e$, to find the model with minimum misfit to the observed topography. Figure \ref{fig:EXFeature} shows an example profile (feature $\#7$) fitted using the numerical method. Figure \ref{fig:EXFeature}a shows the best fit model (orange) compared against the observed topography (blue); Figure \ref{fig:EXFeature}b shows how the misfit varies with elastic thickness for our models. In this case, the best fit $T_e$ is $0.31~\rm km$ with a range of 0.20 to 0.48 km.
More details on the misfit calculation are given below (Section \ref{uncertainty}).

\subsection{Analytical Method}
\label{analytical}

For a feature modeled using the analytical method, we construct a universal flexural profile (UFP) to match the observed topography. This flexural profile models the bending of a plate under a load applied to the end of the plate \citep{turcotte_geodynamics_2002}:

\be
x' = \frac{x - x_0}{x_b - x_0}
\ee

\be
W(x) = w_b \cdot \sqrt{2} \cdot e^{\pi/4} \cdot e^{-(\pi/4) x'} \cdot \sin{\left(\frac{x'\pi}{4}\right)}
\ee
Here $x_b$ is the location of the bulge (the location of maximum elevation); $w_b$ is the height of the bulge above the elevation at the tail end of the profile;
$W(x)$ is the flexural profile (UFP); and $x_0$ is the location of the zero-crossing (where the elevation matches the elevation at the end of the profile).

We vary the parameters $x_b$, $w_b$, and $x_0$ to determine the minimum misfit between the observed topography and the UFP model. With $x_0$ and $x_b$ specified, the UFP uniquely determines the flexural parameter and thus the elastic thickness of the ice shell. We calculate the flexural parameter from the distance between the bulge and the zero-crossing, as described by the following equation from \citep{turcotte_geodynamics_2002}:
\be
x_b - x_0 = \frac{\pi}{4} \cdot \alpha
\ee

Then, using equation 2, the effective elastic thickness can be calculated from the flexural parameter.

\begin{figure}
    \centering
    \includegraphics[width=1.0\textwidth]{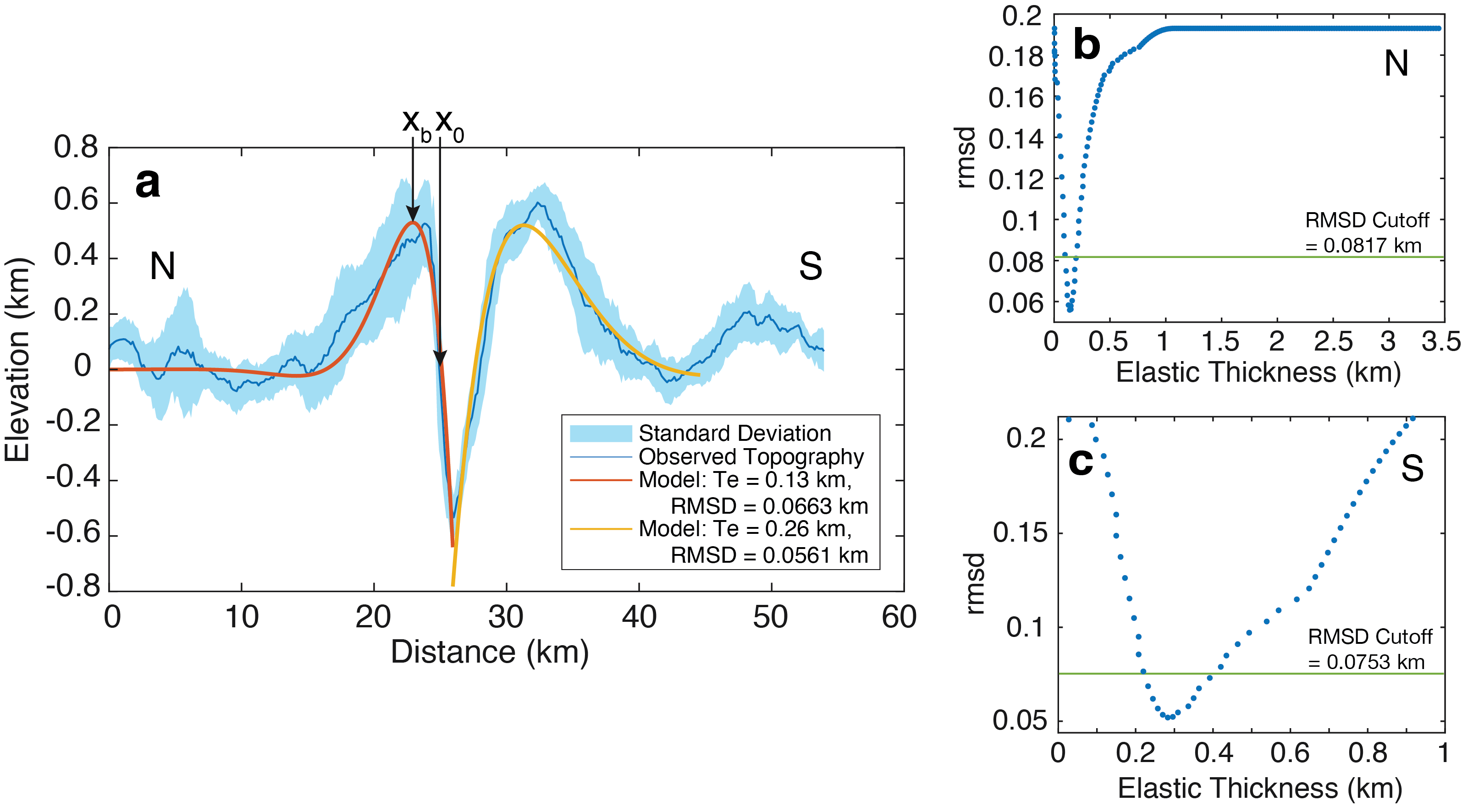}
    \caption{a) Best fit flexural model (orange, yellow) plotted with the averaged topography (blue) for feature $\#10$ (Figure \ref{fig:FeatureMap}, Table \ref{tab:feature tab}). The blue shaded region represents the standard deviation of the topographic profiles. This model is constructed using the analytical method, assuming a universal flexural profile with $x_0$, $x_b$, $w_b$ of $25.1~\rm km, 22.9~\rm km, 0.53~\rm km$ for the northern side and $27.6~\rm km, 31.3~\rm km, 0.52~\rm km$ for the southern side. The best fit model indicates an elastic thickness of 0.13 km on the northern side and 0.26 km on the southern side. 
    b) Misfit of our model as a function of elastic thickness, for the northern side of the feature (orange). The minimum misfit has an RMS deviation of 0.066 km, with a $95\%$ RMSD cutoff value of 0.0817 km, giving an estimated $T_e$ range of 0.11-0.2 km.
    c) Misfit of our model as a function of elastic thickness, for the southern side of the feature (yellow). The minimum misfit has an RMS deviation of 0.056 km, with a $95\%$ RMSD cutoff value of 0.0753 km, giving an estimated $T_e$ range of 0.23-0.39 km.}
    \label{fig:EXFeatureAnalyt}
\end{figure}

Figure \ref{fig:EXFeatureAnalyt} shows an example profile (feature $\#10$) fitted using the analytical method. The best fit models in orange and yellow are plotted against the observed topography, with an independent fit for either side (N,S) of the feature. The locations of the bulge, $x_b$, and the location of the zero-crossing, $x_0$, are indicated for the northern side of the profile. For this feature, the $T_e$ values are 0.13 km with a range of 0.11-0.2 km for the northern side. For the southern side, the best fit $T_e$ is 0.26 km with a range of 0.23-0.39 km.

\subsection{Uncertainty}
\label{uncertainty}

We provide quantification of uncertainty in the elastic thickness estimates as outputs of our flexural models, to determine the range of allowable elastic thicknesses. To quantify the degree of fit between the observations and our model, we used methods from \citep{conrad_2019}. We calculated the root mean square deviation (RMSD) of the model from the observations to represent the degree of misfit.

\be
RMSD = {\left( \frac{1}{N} \sum^{N}_{i=1} {(h_{i,o} - h_{i,m})}^{2} \right)}^{1/2}
\ee
Here, $N$ is the number of points in the observed profile and $h$ is the topography of the observation $(o)$ and model $(m)$. We calculate the misfit for each set of load specifications for the numerical method and over all elastic thicknesses (range of $0-5~\rm km$ in $0.01 ~\rm km$ increments). Similarly, for the analytical method we calculate the misfit for each set of input parameters $(x_b,w_b)$. The minimum misfit determines the best fit model and the corresponding best fit elastic thickness, $T_e$.

To quantify the uncertainty in elastic thickness values, we constructed synthetic topography to match the observed topographic profile and calculated the RMSD (misfit) between the synthetic topography and our best fit model. To construct this synthetic topography, we started with the best fit theoretical model and then added periodic oscillations constructed with an amplitude matching the actual topography outside the feature.
We varied wavelength and amplitude based on the topography outside the feature, determined visually. The amplitude is estimated from the RMSD between this topography and a horizontal line.
Values ranged from $0.05 - 0.2~\rm km$ for amplitude and $1.5 - 30~\rm km$ for wavelength.

\begin{figure}
    \centering
    \includegraphics[width=0.8\textwidth]{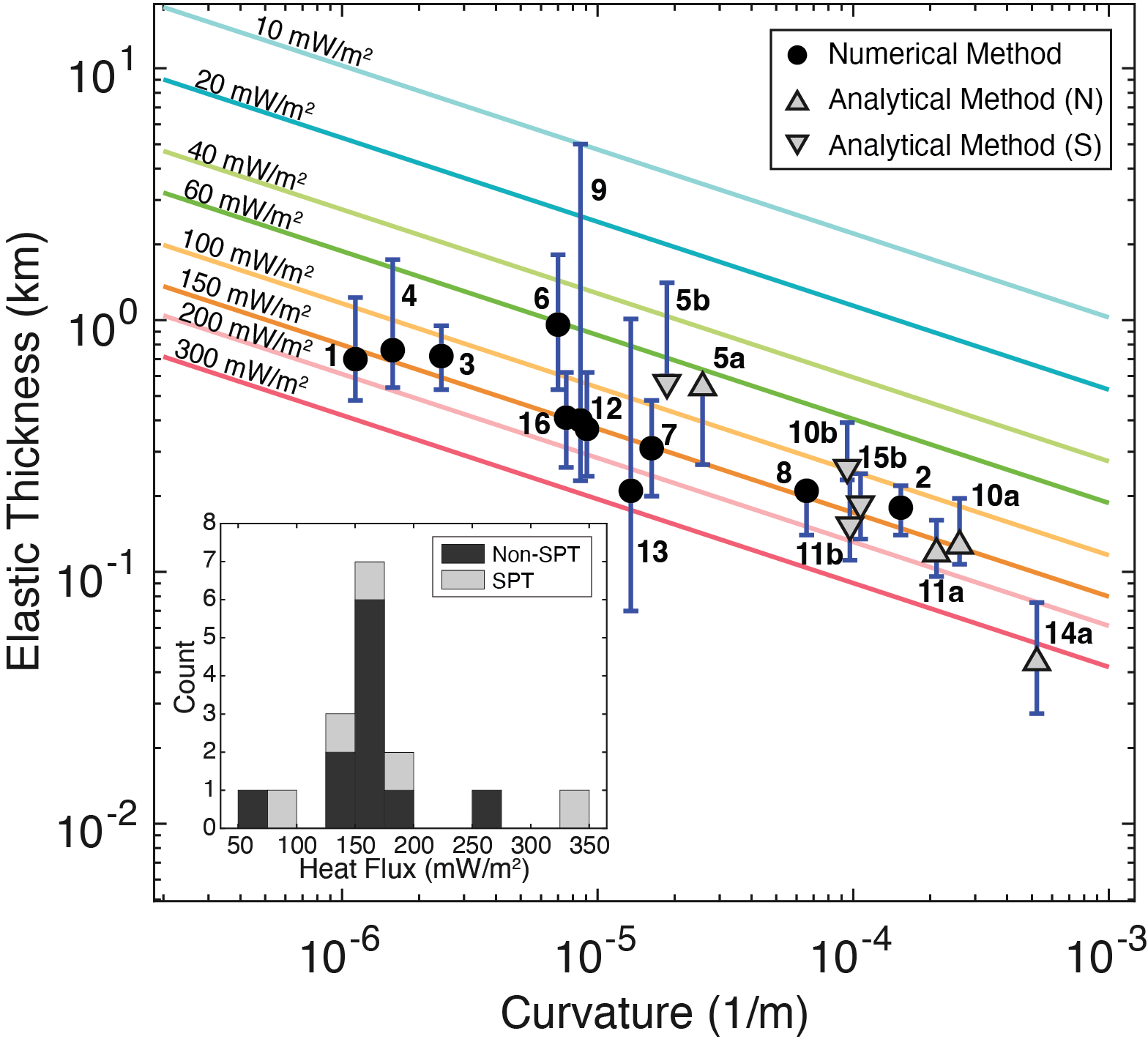}
    \caption{The colored lines show the relationship between elastic thickness, curvature, and heat flux based on the yield strength envelope (YSE) technique \citep{mcnutt_1984} (Section \ref{fitting}). The points represent each feature we analyzed (Figure \ref{fig:FeatureMap}, Table \ref{tab:feature tab}). Filled circles indicate features analyzed using the numerical method, gray triangles indicate features analyzed using the analytical method (upwards = Northern side; downwards = Southern side). Error bars on elastic thickness estimates are calculated using uncertainty methods detailed in Section \ref{uncertainty}. The histogram represents the distribution of paleo-heat fluxes, with light gray indicating SPT margin features analyzed using the analytical method and black indicating features elsewhere that were analyzed using the numerical method. The resulting heat fluxes are in the range of $60\ -\geq350\ \rm mWm^{-2}$, but most are clustered around $125-200~\rm mWm^{-2}$.}
    \label{fig:YSE}
\end{figure}

We then repeated this process to generate 5000 different synthetic noisy topographies to generate a distribution of RMSD values representing misfit between synthetic profiles and noise-free profiles.
We took the RMSD encompassing $95\%$ of our synthetic solutions as the cutoff value for a reasonable misfit to the actual topography (Figure \ref{fig:EXFeature}b, Figure \ref{fig:EXFeatureAnalyt}b,c).
The error bars on our elastic thickness $(T_e)$ in Figure \ref{fig:YSE} represent the range of elastic thickness estimates whose models have a misfit (RMSD) that falls below the critical misfit value.

\subsection{Converting elastic thickness to heat flux}
\label{fitting}

For both methods, the flexural model also provides the radius of curvature. We calculate the curvature at the position where the maximum deflection occurs. We use an approximation for the second derivative to calculate the curvature, $K$.
\be
K = \frac{1}{r_c} = \frac{d^{2}w}{dx^{2}}
\approx \frac{w_{i+1} - 2w_{i} + w_{i-1}}{\Delta x^{2}}
\ee
In this equation, $r_c$ is the radius of curvature in $\rm m$, $\Delta x$ is the spacing between points along the profile, $w$ is the flexural deflection (calculated using the numerical or analytical method), and $i$ is the distance index.

Here we employ the yield-strength envelope (YSE) technique which was introduced by \cite{mcnutt_1984} and is a standard approach for determining heat fluxes in terrestrial settings \citep{watts_isostasy_2001}. Using the YSE approach \citep{mcnutt_1984}, we can determine the relationship between elastic thickness, curvature, and heat flux. It accounts for brittle, elastic, and ductile behavior as opposed to a purely elastic layer (as assumed in our flexural modeling) \citep{watts_isostasy_2001}.
The YSE method assumes a temperature profile of the ice shell and then calculates a stress profile based on the thermal and mechanical properties of the ice shell. We assume a conductive ice shell with a temperature-dependent thermal conductivity, $k_{cond}={651}/{T}$ \citep{petrenko_physics_1999}.
To determine yield stresses, we assume a strain rate of $10^{-13}~\rm s^{-1}$, grain size of $1~\rm mm$, friction coefficient for ice of 0.6 \citep{beeman_friction_1988}, and use the basal-slip accommodated grain boundary sliding rheology from \cite{goldsby_ice_2001}.
More details on this method are included in the Appendix.

\section{Results}
\label{results}

\begin{sidewaystable}
\begin{tabular}{ccccccccccc}
    \toprule
      &   & Ridge or & Lat & Lon & Curvature & RMSD  & Best Fit $T_e$   & $T_e$ Range  & Heat Flux & Terrain Age \\
    Feature $\#$ & Method & Trough? & $(^{\circ}\rm N)$ & $(^{\circ}\rm E)$ & $\rm(m^{-1})$ & (km)  & (km)   & (km)  & $\rm(mWm^{-2})$ & (Gyr) \\  \midrule
    1       & N      & T    & 19.585      & 14.682      & $-1.127*10^{-6}$       & 0.041 & 0.7  & 0.48-1.23 & 100-250         & 4.6         \\
    2       & N      & R    & -1.493      & 22.952      & $1.534*10^{-4}$        & 0.132 & 0.18 & 0.14-0.22 & 100-150         & 3.9         \\  
    3       & N      & T    & -10.298     & 37.279      & $-2.445*10^{-6}$       & 0.073 & 0.72 & 0.53-0.95 & 100-175         & 2.6         \\  
    4       & N      & T    & -5.854      & 53.848      & $-1.577*10^{-6}$       & 0.072 & 0.76 & 0.54-1.74 & 60-200          & 1.8         \\  
    5a      & A      & T    & -56.950     & 24.582      & $-2.574*10^{-5}$       & 0.049 & 0.55 & 0.27-0.56 & 60-150          & 0.3         \\  
    5b      & A      & T    & -56.950     & 24.582      & $-1.867*10^{-5}$       & 0.041 & 0.56 & 0.6-1.41  & 30-80           & 0.3         \\  
    6       & N      & T    & -26.677     & 72.336      & $-6.992*10^{-6}$       & 0.134 & 0.96 & 0.53-1.82 & 30-125          & 2.0         \\  
    7       & N      & R    & 31.003      & 101.500     & $1.626*10^{-5}$        & 0.047 & 0.31 & 0.2-0.48  & 100-250         & 4.1         \\  
    8       & N      & R    & 5.223       & 129.446     & $6.569*10^{-5}$      & 0.261 & 0.21 & 0.14-0.22 & 140-200         & 3.9         \\  
    9       & N      & T    & 10.451      & 170.800     & $-8.567*10^{-6}$    & 0.112 & 0.4  & 0.23-5    & 10-250          & 4.6         \\  
    10a     & A      & T    & -50.528     & 270.358     & $-2.607*10^{-4}$       & 0.066 & 0.13 & 0.11-0.2  & 100-175         & 0.4         \\  
    10b     & A      & T    & -50.528     & 270.358     & $-9.469*10^{-5}$       & 0.056 & 0.26 & 0.23-0.39 & 60-125          & 0.4         \\  
    11a     & A      & T    & -52.181     & 279.906     & $-2.116*10^{-4}$       & 0.061 & 0.12 & 0.1-0.16  & 125-200         & 0.4         \\  
    11b     & A      & T    & -52.181     & 279.906     & $-9.704*10^{-5}$      & 0.044 & 0.15 & 0.11-0.27 & 100-250         & 0.4         \\  
    12      & N      & R    & -23.515     & 300.701     & $9.078*10^{-6}$        & 0.035 & 0.37 & 0.24-0.62 & 100-250         & 1.1         \\  
    13      & N      & T    & -27.453     & 310.225     & $-1.348*10^{-5}$       & 0.050 & 0.21 & 0.07-1.01 & 50-400          & 0.7         \\  
    14a     & A      & T    & -53.220     & 313.064     & $-5.228*10^{-4}$       & 0.039 & 0.04 & 0.03-0.08 & 200-400         & 4.0         \\  
    15a     & A      & T    & -51.289     & 351.530     & $4.136*10^{-3}$        & 0.031 & 0.01 & 0.01-0.09 & -               & 0.5         \\  
    15b     & A      & T    & -51.289     & 351.530     & $-1.067*10^{-4}$      & 0.045 & 0.18 & 0.14-0.25 & 100-200         & 0.5         \\  
    16      & N      & T    & 7.738       & 352.795     & $-7.528*10^{-6}$       & 0.041 & 0.41 & 0.26-0.62 & 100-250         & 4.6         \\ \bottomrule
\end{tabular}
\caption{Table of features analyzed. Feature numbers correspond to Figure \ref{fig:FeatureMap}. Method refers to analytical (A) versus numerical (N). For the analytical method, (a) = Northern side of feature; (b) = Southern side of feature. Curvature, RMSD, and best fit $T_e$ are outputs of the best fit model. The ranges of paleo-heat fluxes are determined from Figure \ref{fig:YSE}, representing the widest range of acceptable heat fluxes. Terrain age is estimated from local crater density (described more in Section \ref{craters} and determined from Figure \ref{fig:cratering}).}
\label{tab:feature tab}
\end{sidewaystable}

We find a range of best fit elastic thicknesses from $0.01\ \rm km$ to $0.96\ \rm km$, with corresponding heat fluxes at the time of deformation ranging from $60\ \rm mWm^{-2}$ to $\geq 350\ \rm mWm^{-2}$ (Figure \ref{fig:YSE}, Table \ref{tab:feature tab}). Most of our paleo-heat fluxes are clustered around $125-200~\rm mWm^{-2}$, with a few outliers having average values that are lower (e.g. feature $\#6$) or higher (e.g. feature $\#14a$). 
Figure \ref{fig:YSE} shows the relationship between elastic thickness, curvature of deformation, and heat flux. Our results are plotted as points with error bars indicating the acceptable range of $T_e$ values (methods detailed in Section \ref{uncertainty}). The range of paleo-heat fluxes reported in Table \ref{tab:feature tab} for each feature is based on the range of acceptable elastic thicknesses.
The elastic thickness estimates we obtain record the minimum $T_e$ since the load was emplaced, which corresponds to the highest heat flux the feature experienced.

Feature $\#9$, located in the non-polar cratered terrains, warrants further discussion. It is the only feature we analyzed that is located in heavily cratered terrain. This feature has a shortened profile because it is enclosed on either side by craters, making the fitting process difficult and resulting in a large uncertainty for elastic thickness. We only report the lower bound elastic thicknesses of $0.23~\rm km$ for this feature, which implies an upper-bound heat flux of $250~\rm mWm^{-2}$ (Table \ref{tab:feature tab}).

We exclude feature $\#15a$ from our analysis because there is minimal flexural deformation at the northern side of this feature, which results in unrealistic elastic thickness and paleo-heat flux estimates.

\subsection{SPT Margins}
\label{margins}

The SPT margins are characterized by the intersection of the younger surface of the SPT with the more heavily cratered terrain outside the SPT and the presence of curvilinear features that encircle the SPT \citep{patterson_icymoons_2018,crow-willard_mapping_2015}. The margins are geologically interesting, as it has been suggested that the ice shell could be experiencing compressional and extensional stresses on either side of the SPT \citep{yin_spt_2015}.

Features $\#$5,10,11,14, and 15 are located in the SPT margins \citep{crow-willard_mapping_2015,yin_spt_2015} (Figure \ref{fig:FeatureMap}, Table \ref{tab:feature tab}). Most of these features are located in the leading hemisphere at latitudes south of $50^\circ$ S.
For these features in the SPT margins, we find paleo-heat fluxes ranging from $60-350\ \rm mWm^{-2}$. 
The paleo-heat fluxes we find are broadly consistent with the present-day heat fluxes measured inside the SPT \citep{howett_2011}.

We fit both sides of the SPT margin features independently using the analytical method, finding two separate elastic thickness estimates and similarly two different paleo-heat flux estimates on each side of the margin. Comparing the northern side (a) with the southern side (b) (Figure \ref{fig:YSE}), we find that there is no clear pattern in either the elastic thickness estimates or the heat fluxes.
For example, features $\#$5 and 10 have a lower elastic thickness and a higher heat flux on the northern side. However, feature $\#11$ has a higher elastic thickness and a lower heat flux on the northern side (Table \ref{tab:feature tab}).

The average heat flux in the SPT margins is similar to heat fluxes elsewhere (Figure \ref{fig:YSE}), with both regions having the majority of their heat fluxes clustered around $125-200~\rm mWm^{-2}$. We find that the margins do not have a significant difference in elastic thickness or heat flux on either side. We cannot make strong conclusions about any heat flux difference along the SPT margins, but it is a region that warrants further study due to the connection between the region's geologic and thermal history.

\section{Discussion}
\label{discussion}

\subsection{Comparison with present-day heat flux estimates}
\label{present hfs}

Present-day infrared measurements within the SPT place heat fluxes in the range of $100-500~\rm mWm^{-2}$ \citep{howett_2011}.
Our highest estimates of paleo-heat fluxes, of $\geq350\ \rm mWm^{-2}$ (e.g. features $\#13,14$), are of similar magnitude to the present-day heat flux measurements at the SPT.
Similar paleo and present-day heat fluxes could be an indication that the peak heat flux produced by Enceladus has not changed over time (Section \ref{craters}).
Furthermore, since $\#13$ is located in the southern mid-latitudes, not close to the SPT (the current location of peak heating), the locus of maximum heating may have moved over time.

Infrared measurements of the northern polar region record a value of around $50~\rm mWm^{-2}$ \citep{miles_npheat_2025}.  
Due to the lack of chasmata and dorsae in the northern latitudes (above $70^\circ$N), we are unable to provide a direct comparison between paleo-heat fluxes and present-day measurements of heating at the north polar region.
However, our estimated paleo-heat fluxes in the mid-latitudes and equatorial regions (e.g. feature $\#6$ with $30-125~\rm mWm^{-2}$) are comparable to measured heat flux at the north polar region \citep{miles_npheat_2025}.

\begin{figure}
    \centering
    \includegraphics[width=0.9\textwidth]{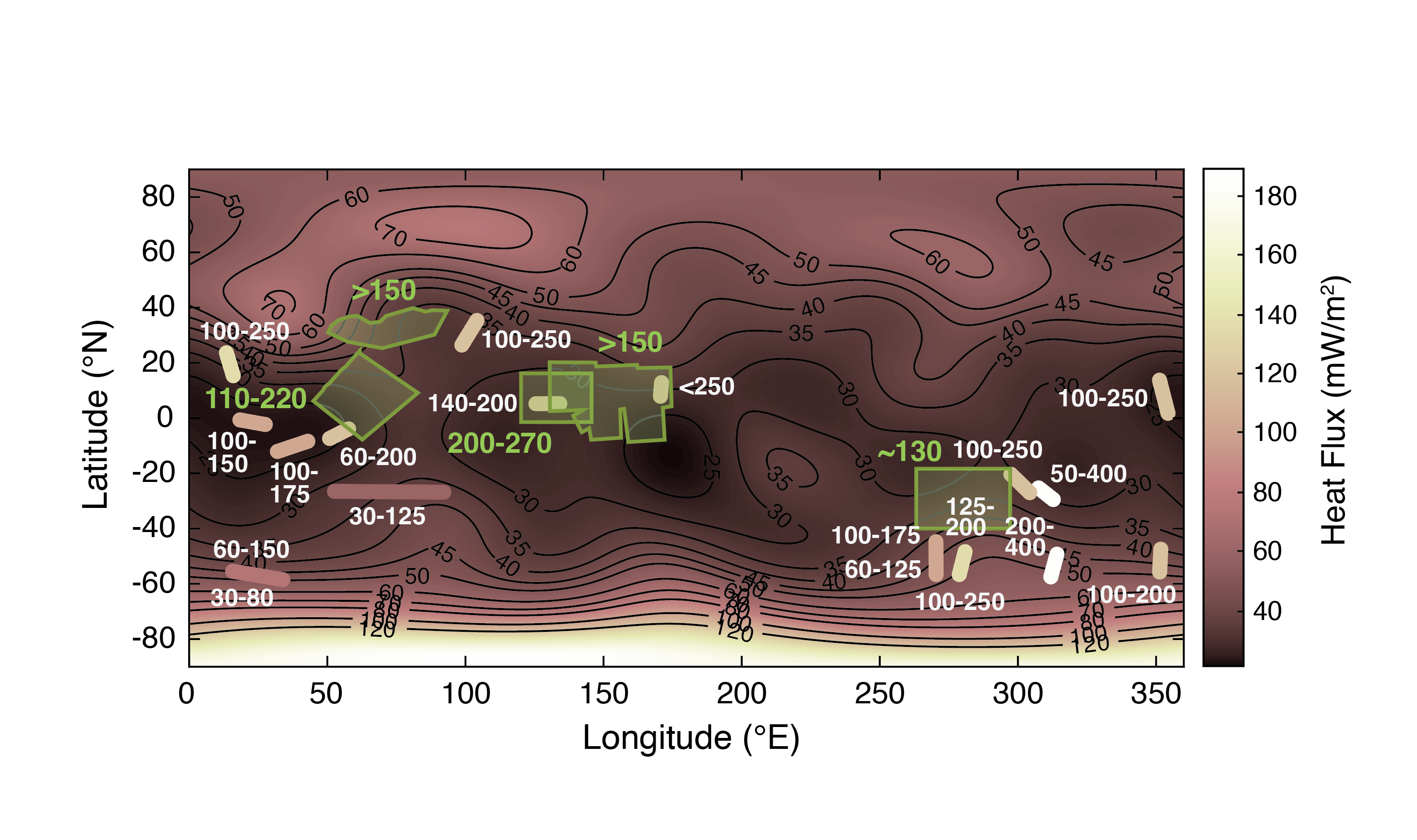}
    \caption{Global map of Enceladus with color and contours indicating the estimated heat flux (in $\rm mWm^{-2}$) calculated from maps of shell thickness \citep{hemingway_mittal_2019}. Our results (Table \ref{tab:feature tab}) are plotted as lines using the same color scale as the background.
    Green labeled boxes show other estimates \citep{giese_2008,bland_crater_relaxation_2012,bland_2007,leonard_2021} of paleo-heat fluxes using similar methods.}
    \label{fig:HFMap}
\end{figure}

Based on inferred shell thickness variations calculated from topography and gravity and assuming a conductive ice shell, the local present-day heat flux can be derived \citep{hemingway_mittal_2019}.
Figure \ref{fig:HFMap} shows these present-day heat flux estimates for Enceladus in the background color map and contour lines. Based on the shell thickness variations, the SPT has the thinnest ice, and consequently the highest present-day conductive heat fluxes ($>120~\rm mWm^{-2}$) \citep{hemingway_mittal_2019}.

\begin{figure}
    \centering
    \includegraphics[width=0.8\textwidth]{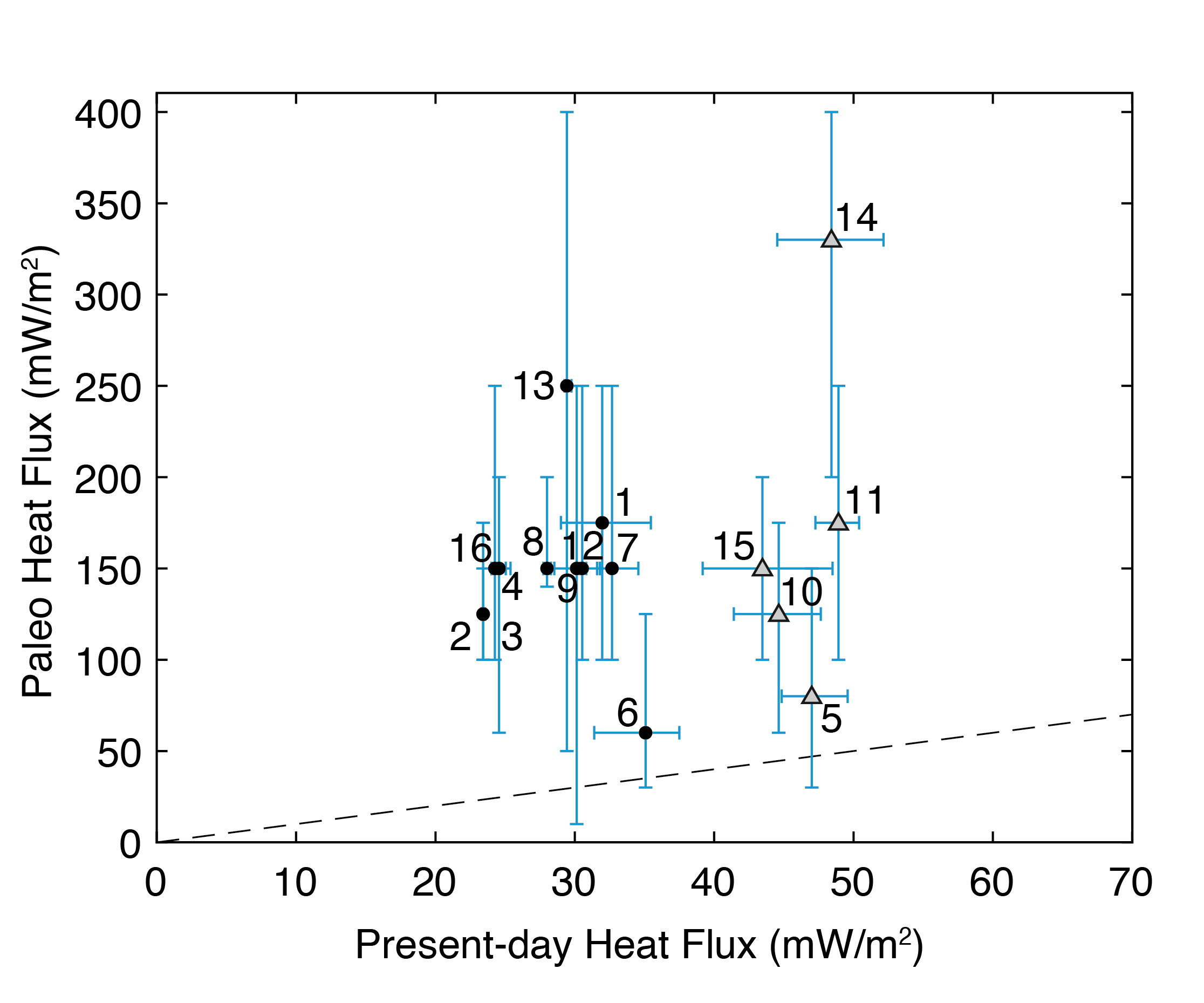}
    \caption{Comparison of paleo-heat flux estimates with present-day values in the same location derived from shell thickness models \citep{hemingway_mittal_2019}. Each point represents a feature we analyzed, (Figure 1, Table 1). Filled circles indicate features analyzed using the numerical method, gray triangles indicate features analyzed using the analytical method. The dashed line indicates a 1:1 linear relationship between paleo- and present-day heat fluxes. Error bars in paleo-heat flux estimates are derived from our uncertainty estimations (Section \ref{uncertainty}).}
    \label{fig:HFCompare}
\end{figure}

The paleo-heat fluxes we find are greater than the present-day heat fluxes inferred from shell thickness models in the same locations \citep{hemingway_mittal_2019}. In Figure \ref{fig:HFCompare}, we compare our paleo-heat flux estimates with the present-day values derived by \cite{hemingway_mittal_2019} for the same area. All our features plot above the dashed line representing equal present and paleo-heat fluxes (Figure \ref{fig:HFCompare}), which means that our paleo-heat flux estimates are greater than present-day heat fluxes at the same locations. 
For example, feature $\#2$, located in the trailing hemisphere near the equator, has an estimated ancient heat flux of $100-150\ \rm mWm^{-2}$ and a present day heat flux of $25\ \rm mWm^{-2}$.

Greater paleo-heat fluxes than present-day values could imply a cooling and thickening ice shell. Our results indicate that this cooling and ice shell thickening could be occurring outside of the SPT at present, in the equatorial and mid-latitude regions of the trailing and leading hemisphere terrain. Localized episodic heating followed by cooling and ice shell thickening is suggested by \cite{crow-willard_mapping_2015,ojakangas_episodic_1986}. 
Taking feature $\#2$ as an example, heat flux has declined from approximately $100~\rm mWm^{-2}$ to $25~\rm mWm^{-2}$ at the present day. The corresponding change in shell thickness would be from 8 to 32 km. Neglecting other sources of heat, conductive shell thickening of this magnitude would only take $\approx 1$ Myr, although the presence of heat sources would prolong the cooling timescale.
A potential explanation for the declining heat flux is that the pattern of tidal heating has shifted over time, with the peak heating located at the SPT currently.
Transient thermal events at the SPT have been suggested by \cite{yin_spt_2015}, as a potential requirement for the gravitational spreading model for SPT formation.

\subsection{Comparison with other ancient heat flux estimates}
\label{past hfs}

In the equatorial regions of the trailing hemisphere (Sarandib and Diyar Planitia), \cite{bland_2007} found heat fluxes ranging from $110-220\ \rm mWm^{-2}$ based on analyzing ridges and troughs potentially generated from unstable extension of the lithosphere. Our paleo-heat flux estimates in this area are similar at $60-200\ \rm mWm^{-2}$ (feature $\#4$, Table \ref{tab:feature tab}).
\cite{giese_2008} modeled plate flexure and found heat fluxes ranging from $200-270\ \rm mWm^{-2}$; assuming higher surface porosity of the ice shell reduced their heat flux estimates to $45-60\ \rm mWm^{-2}$.
Our paleo-heat fluxes in this area are estimated at $140-200\ \rm mWm^{-2}$ (feature $\#8$, Table \ref{tab:feature tab}).
In the northern latitudes of Enceladus, \cite{bland_crater_relaxation_2012} quantified the relaxation of shallow craters, finding heat fluxes in excess of $150\ \rm mWm^{-2}$. Our estimates of $100-250\ \rm mWm^{-2}$ are consistent with these results (feature $\#7$, Table \ref{tab:feature tab}).

There are few estimates of paleo-heat fluxes in the leading hemisphere (180 to $360^\circ$ E), but one study finds heat fluxes of $130\ \rm mWm^{-2}$ under the assumption that tidal stresses dominate heating \citep{leonard_2021}. Our estimates of $50-\geq350\ \rm mWm^{-2}$ in the leading hemisphere (e.g. features $\#12,13,16$) are consistent with these findings.

\subsection{Lack of correlation between terrain age and paleo-heat flux}
\label{craters}

We investigated the relationship between the heat flux estimates and the crater density as a proxy for terrain age, which could reveal time-dependence of heating (e.g. higher heat fluxes might be found in younger regions with lower crater density). The crater density of Enceladus' surface was mapped by \cite{kirchoff_crater_2009}. We use their global crater database in combination with the cratering rate model suggested by \cite{zahnle_cratering_2003} to derive an estimated terrain age for each of our features.

We exclude craters with diameters smaller than $1~\rm km$, as \textit{Cassini} images of Enceladus' surface have a resolution of $>0.1~\rm km$/pixel, which makes crater identification below that scale difficult. This also reduces the effect of counting secondary craters, which would artificially inflate crater counts and result in older terrain ages \citep{kirchoff_crater_2009}.
There are relatively fewer small craters (of diameters 0.6 km to 2 km) on Enceladus as compared to the other Saturnian satellites, which has been suggested to be a result of geologic activity (plume material or E-ring material covering smaller craters in the equatorial regions) and viscous relaxation removing larger craters (diameters greater than 6 km) \citep{kirchoff_crater_2009}.

We take a local area around the features (within a latitude and longitude box) and count the craters inside the area, normalizing to the area. The areas range from $(0.3-5.2) *10^4~\rm km^2$, equivalent to $0.44\%$ to $6.5\%$ of the surface area of Enceladus. Box areas are depicted in the Appendix in Figure \ref{fig:Crater_map}.
Once crater density estimates are obtained, we apply the logarithmic model, Z03 Case A \citep{zahnle_cratering_2003}, relating the crater density on a surface to the terrain age. This model is specific to the outer solar system, assuming a lack of smaller impactors in order to match the expected cratering rates for bodies like Europa and Ganymede. This model is consistent with the crater size distribution observed on Pluto and Charon \citep{singer_impact_2019}.

Our results suggest a range of surface ages from 0.3 Gyr - 4.6 Gyr, with no clear correlation between the magnitude of paleo-heat flux estimates and terrain age. Estimations of absolute terrain age are uncertain due to uncertainty in the impact flux, over-counting of secondary craters, and removal of craters due to potential resurfacing or crater relaxation; nonetheless relative surface ages are likely correct. From Monte Carlo simulations of the expected variation in crater density as a function of box area, we find that the expected variation in crater density is at most $\sim 30\%$, which results in variation of terrain age up to $\sim 30\%$. The terrain ages we obtain are calculated using a globally averaged impactor flux model, which does not account for hemispheric differences in impactor flux.

\begin{figure}
    \centering
    \includegraphics[width=0.8\textwidth]{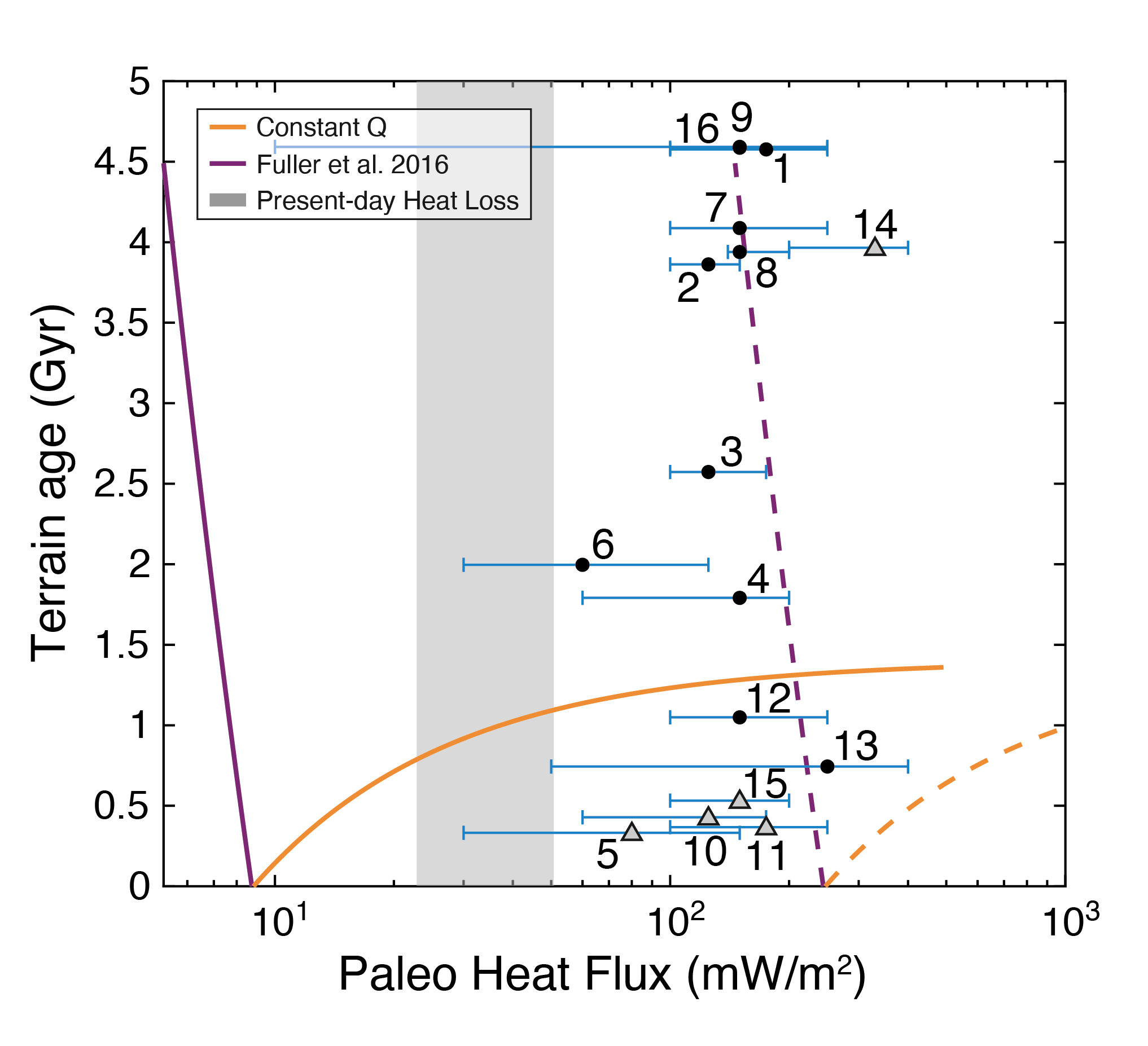}
    \caption{Estimated terrain age, calculated using model Z03 (Case A) \citep{zahnle_cratering_2003} and maps of cratering across Enceladus \citep{kirchoff_crater_2009}, plotted against our estimated paleo-heat fluxes. Blue error bars represent the ranges of uncertainty in the heat flux estimates. Each point represents a feature, with numbers corresponding to Figure \ref{fig:FeatureMap}, Table \ref{tab:feature tab}. Filled circles indicate features analyzed using the numerical method; gray triangles indicate features analyzed using the analytical method.
    Equilibrium tidal heating rate models at Enceladus are plotted in orange (constant tidal quality factor, \textit{Q}, of Saturn) and in purple (resonance locking, \citep{fuller_resonance_2016}) replicated from \citep{nimmo_dissipation_2023}. Here we take $Q = 3100$ and the internal evolution timescale of Saturn to be $t_\alpha = 13~\rm Gyr$ \citep{nimmo_dissipation_2023}.
    Due to uncertainty in the distribution of tidal torques between Dione and Enceladus, tidal heat production rates range from the solid to the dashed lines (see Section 
    \ref{craters}). The grey shaded region represents the present-day global heat loss estimates (18-40 GW) \citep{hemingway_mittal_2019,park_enceladus_shape_2024}, corresponding to heat fluxes of $23-50~\rm mWm^{-2}$ at Enceladus as derived from globally averaged shell thickness models.}
    \label{fig:cratering}
\end{figure}

Figure \ref{fig:cratering} shows the estimated terrain age plotted against our paleo-heat flux estimates. We find heat fluxes $>100~\rm mWm^{-2}$ present at both younger (<1 Gyr) and older (4 Gyr) terrains (e.g. features $\#15,14$). The lack of correlation between magnitude of paleo-heat fluxes and terrain age implies that the peak heat flux on Enceladus has not changed over time. Taking the impact-derived ages at face value, Enceladus has generated, at least locally, heat fluxes of roughly $100~\rm mWm^{-2}$ over billions of years. This has implications for its orbital evolution \citep{nimmo_dissipation_2023}.

The present-day cause of Enceladus's tidal heating is its 2:1 mean motion resonance (MMR) with Dione \citep{meyer_tidal_evol_2008,meyer_tidal_heat_2007}. The resonance maintains Enceladus's eccentricity and tidal heating; the ultimate source of the energy is rotational kinetic energy of Saturn, some of which is transferred to the satellites via tidal torques. The rate of energy transfer depends on the tidal dissipation factor, \textit{Q}, of Saturn e.g. \citep{nimmo_icymoons_2018}. In general, \textit{Q} can be both frequency and time-dependent \citep{fuller_resonance_2016,lainey_tidal_dissipation_2012}. 

If the satellites in a MMR have eccentricities that are constant, the equilibrium tidal heat production can then be calculated without needing to know the satellites' deformative properties \citep{meyer_tidal_heat_2007}. In the case of a constant \textit{Q} of Saturn, the equilibrium heat fluxes are much higher at early times, because the satellites are closer to Saturn. As the satellites migrate outwards due to dissipation in Saturn, tidal heating decreases due to its dependence on the orbital distance of the satellite.

In an alternate formulation ("resonance locking" \citep{fuller_resonance_2016}), the \textit{Q} of Saturn is time-variable and is controlled by the rate at which Saturn's internal structure evolves. In this picture, the equilibrium tidal heat production increases only moderately compared with earlier times \citep{nimmo_dissipation_2023}.

Inspection of Figure \ref{fig:cratering} shows that the "resonance locking" model is more consistent with our results than the constant-\textit{Q} model. Although the age of the Enceladus:Dione resonance is not known, our results suggest that it is ancient (subject to the caveat that the impact flux is uncertain).

A complication with the theory of equilibrium tidal heating \citep{meyer_tidal_heat_2007} is that the answers vary depending on whether both Dione and Enceladus are being torqued by Saturn (solid line), or only Enceladus is (dashed line) \citep{nimmo_dissipation_2023}. In the former case, the predicted heat fluxes are lower. However, for that to be true both Enceladus and Dione would have to be exciting a resonant response inside Saturn, which is improbable. More likely is that only Enceladus is doing so, and indeed this is what the results in Figure \ref{fig:cratering} suggest.

A further complication is that Dione and Enceladus can experience periodic tidal heating (though the long-term average must conform to the equilibrium value) \citep{ojakangas_episodic_1986,meyer_episodic_2008}. Alternatively, Dione may have experienced an eccentricity-exciting resonance prior to its entry into the present-day resonance with Enceladus \citep{meyer_tidal_evol_2008}. Either of these cases raise the possibility that Dione could have been heated in the past. If all the equilibrium heating were concentrated in Dione, the mean surface heat flux would be about $50~\rm mWm^{-2}$. This is consistent with measurements of flexure on Dione that place surface heat fluxes in the range $25-60~\rm mWm^{-2}$ \citep{hammond_dione_flexure_2013}.

\subsection{Caveats and Uncertainties}
\label{caveats}

An important caveat is that comparison between global tidal heat production models and local paleo-heat fluxes is not a consistent metric. Inherent to this comparison is the assumption that local heat fluxes are a representation of the equilibrium global heat flux, which is not necessarily true.

Uncertainty in impactor rates in the Saturnian system could influence absolute terrain age estimates, which constrains our estimates as upper bounds.
Orbital instabilities or satellite collisions could form planetocentric impactors \citep{ferguson_elliptical_2022} thereby inflating crater density.
If the ages of the Saturnian satellites were significantly younger (potentially as young as a few $100~\rm Myr$) \citep{cuk_dynamical_2016,cuk_tidal_evol_2023,cuk_instability_2026,wisdom_rings_2022}, that would also imply that our terrain age estimates are upper bounds. However, the uncertainty in absolute terrain ages do not change the implications of our results. Even if the surface was younger than reported here, we still do not find significantly higher paleo-heat fluxes than peak present-day values. Our results would still be more consistent with resonance locking over the constant-\textit{Q} model of tidal heat production.

Some uncertainty in our conclusions arises because of the combination of temporal and spatial variations in heating. Flexural modeling records the highest heat flux since the ice shell deforms. Higher episodes of heating would overwrite the previous degree of deformation, causing the ice shell to not record periods of cooling. This means that our results do not rule out cyclic tidal heating, as suggested by \citep{goldreich_enceladus_2025,meyer_tidal_heat_2007},
but more research would need to be done to determine whether observed geologic features support the predicted tidal heating models.
It is difficult for these methods to disentangle variations in heat flux due to local changes in heating (such as the locus of heating moving spatially to the SPT) and temporal changes in heating (such as variations in the magnitude of tidal heating, either due to eccentricity excitation or damping).

The radial distribution of tidal heating, i.e. how the heat gets dissipated in Enceladus is not well understood. Prevailing theories suppose the majority of dissipation occurs in a water-filled and porous core \citep{roberts_fluffy_core_2015,choblet_heating_2017}. Heating in the ocean or ice shell is likely insufficient \citep{soucek_tidal_2019,hay_dissipation_2019} 
to reproduce the total heat loss at Enceladus of $30-50~\rm GW$.
Determining how heat is dissipated across the moon is a non-trivial problem that has important implications for the longevity of a global subsurface ocean and the mechanics of localized hot spot migration across the surface.

Additional uncertainty in heat flux arises from the choice of constants for the YSE method. We varied the strain rate from $10^{-12}~\rm s^{-1}$ to $10^{-17}~\rm s^{-1}$ and the grain size from 0.1 mm to 1 cm to determine the sensitivity of our results on these variables. At most, these variations changed heat flux estimates by $\sim25-50~\rm mWm^{-2}$, or up to $\sim30\%$. A similar range of uncertainties was obtained by \cite{nimmo_flexure_ganymede_2004} in a YSE study of Ganymede. The effect of changing strain rate and grain size is more significant at higher heat fluxes and lower elastic thicknesses.

\section{Conclusions}
\label{conlusions}

In this work, we present estimates of paleo-heat fluxes across the surface of Enceladus. We find paleo-heat fluxes across Enceladus of $60-\geq350~\rm mWm^{-2}$ that are consistent with other estimates \citep{bland_2007,bland_crater_relaxation_2012,leonard_2021,giese_2008}.

For regions outside the SPT, paleo-heat fluxes are higher than present-day heat fluxes estimated from shell thickness maps in the same locations \citep{hemingway_mittal_2019}. This implies that the ice shell is cooling and thickening outside of the SPT. Both the trailing hemisphere and leading hemisphere terrains experienced high heat fluxes in excess of $100~\rm mWm^{-2}$ in the past, potentially due to an ancient heating episode (either global or local). Prior studies also suggested ancient periods of intense heating on Enceladus \citep{bland_2007,bland_crater_relaxation_2012}.

We find paleo-heat fluxes in equatorial and mid-latitude locations that are comparable to both estimated and measured values of present-day heat fluxes at the SPT \citep{hemingway_mittal_2019,howett_2011}. This suggests that the peak heat flux recorded on Enceladus has not changed significantly over time, but the location of peak heating has. Further investigation into the spatial distribution of paleo-heat fluxes, such as modeling crater relaxation in the Northern hemisphere, would help determine where geologically active regions were located in the past. Future missions to Enceladus with an orbiter or lander would provide improved geodetic and/or seisomological measurements, which could further constrain the present-day ice shell thickness variations across the moon and potential differences in interior structure (\cite{bagheri_structure_2025,dapre_seismic_2025}).

Finally, we compared our paleo-heat fluxes with terrain age estimates and find no correlation between terrain age and paleo-heat flux. This result supports the idea of a near-constant peak heat flux on Enceladus over its history. Our results are supportive of the resonance locking model of tidal heat production \citep{fuller_resonance_2016}. 
If resonance locking is happening, our results would suggest that Enceladus and Dione have been in a long-lived MMR and that Enceladus is experiencing the majority of tidal torques and thus a higher magnitude of tidal heating than Dione. This conclusion supports the idea of a long-lived subsurface ocean on Enceladus.

\bibliographystyle{agufull}
\bibliography{refs_enc}

\section{Appendix}
\label{supplement}

\beginsupplement

\subsection{Crater density correlation with paleo-heat flux}
\label{supplement crater dens}

There is a strong dependence of terrain ages on latitude (excluding feature $\#14$, as discussed below). We find that the oldest terrains and most cratered features are located near the northern latitudes of Enceladus. The latitude dependence of crater density on Enceladus has been noted previously by other studies \citep{kinczyk_cratered_terrains_2024,kirchoff_crater_2009}. It has also been suggested that this latitude dependent crater distribution could be a result of resurfacing from plume ejections at the SPT \citep{kirchoff_crater_2009}.

\begin{figure}
    \centering
    \includegraphics[width=0.9\textwidth]{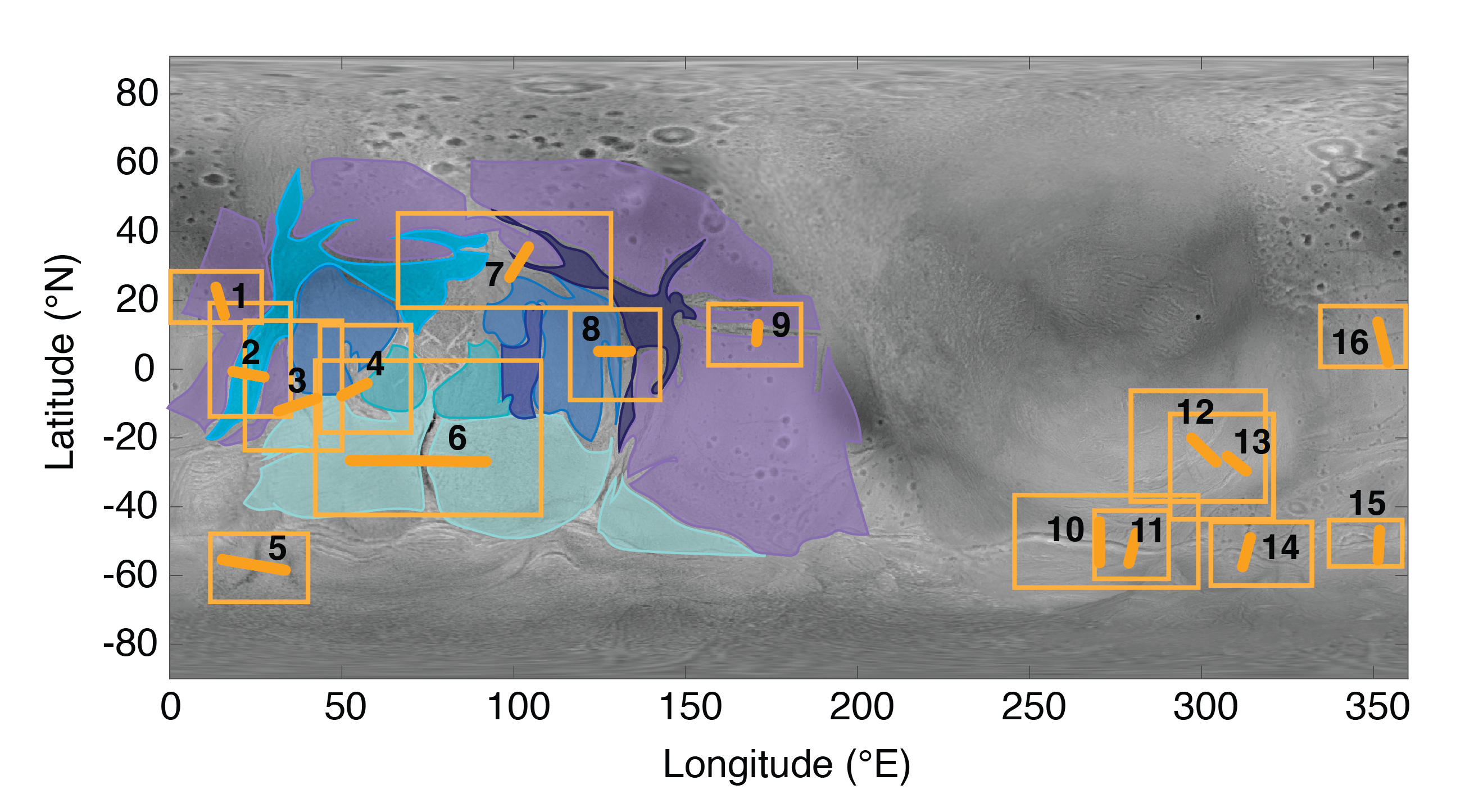}
    \caption{Digital elevation map (DEM) of Enceladus from \cite{schenk_2024} plotted with orange lines representing topographic profiles across chasmata and dorsae. Orange boxes around each feature indicate the latitude and longitude boxes where craters were counted to estimate local terrain ages. Each feature is indicated with a number, as shown in Table \ref{tab:feature tab}. Shaded regions represent geologic units from \cite{kirchoff_crater_2009}, adapted from their Figure 4. The light purple shaded region represents the cratered plains, while the blue shaded regions represent ridged plains.}
    \label{fig:Crater_map}
\end{figure}

The terrain maps that \cite{kirchoff_crater_2009} produced are located in the trailing hemisphere, but their crater database is global. Figure \ref{fig:Crater_map} shows the location of features analyzed here against the mapped regions of \cite{kirchoff_crater_2009}.
Our features $\#1-4,6-9$ reside in the mapped terrain (Figure \ref{fig:Crater_map}) and we can compare terrain age estimates between our method and that of \cite{kirchoff_crater_2009}. Features $\#5,10-16$ are not in the classified terrains as specified by the \cite{kirchoff_crater_2009} maps.

\cite{kirchoff_crater_2009} find absolute terrain ages that are consistent with our estimates for the heavily cratered regions, but younger for less cratered areas. For example features $\#1,9$ reside in the equatorial cratered plains terrain \citep{kirchoff_crater_2009,crow-willard_mapping_2015}, which has an age of 4.4 Gyr. We find an estimated terrain age of $>4.5$ Gyr.
For less cratered regions, such as the ridged plains, \cite{kirchoff_crater_2009} find a terrain age of 1.5 Gyr. For example, feature $\#3$, which our estimates of terrain age place at 2.5 Gyr. This discrepancy could be due to differences in localized crater density surrounding this feature, as our estimated terrain age will bias older if there are regions of higher crater density that are included near the feature. We ran Monte Carlo simulations to determine the expected variation in crater density as a function of box area. We find that the expected variation in crater density for our smallest area is only $\sim30\%$ of the actual value, which means that our estimates of terrain age could vary by $\sim30\%$.

The age of features do not change significantly when craters with a lower confidence level ($<50\%$) as determined by \cite{kirchoff_crater_2009} are excluded. We included all identified craters regardless of confidence level in our calculation of absolute terrain age. Removing lower confidence craters decreases our inferred terrain ages by at most $0.3~\rm Gyr$.

\begin{figure}
    \centering
    \includegraphics[width=0.9\textwidth]{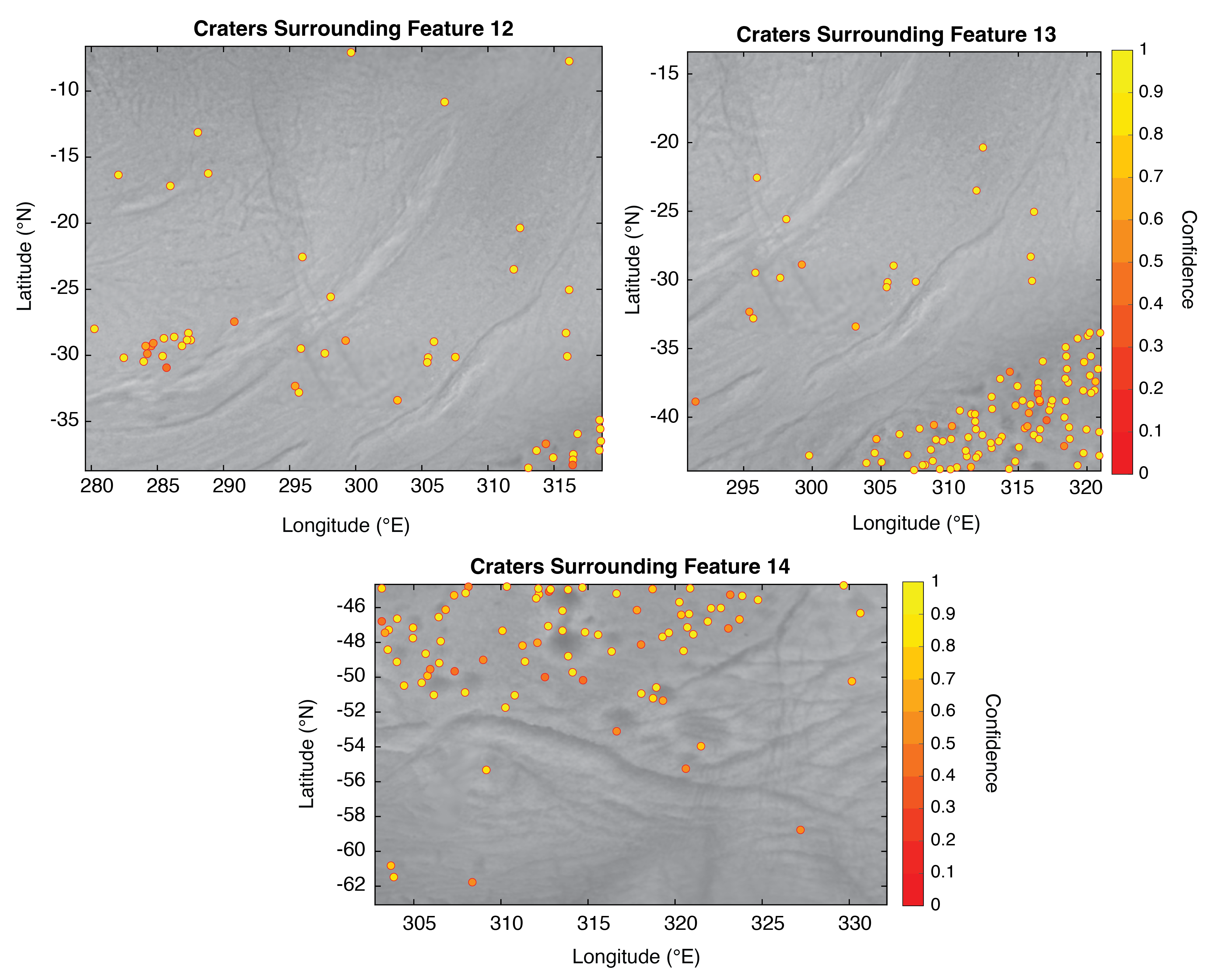}
    \caption{Digital elevation map (DEM) of Enceladus from \cite{schenk_2024} with elevations from +2.3391 km (white) to -2.7330 km (black). Circles represent craters $\geq$ 1km, where color corresponds to the confidence level \citep{kirchoff_crater_2009}.}
    \label{fig:Crater_examples}
\end{figure}

Features $\#12,13$ are potentially contaminated with higher cratering counts from other terrain. These two features are located in the leading hemisphere terrain (LHT) curvilinear region, which is adjascent to cratered plains \citep{crow-willard_mapping_2015}. In the southeast corner of the surrounding area, the terrain changes from being tectonized to being heavily cratered (Figure \ref{fig:Crater_examples}). These craters are excluded to get a better estimate of the age of the local terrain around these two features. The estimated age for these two features reduces from 1.3 Gyr to 1.0 Gyr and 2.9 Gyr to 0.7 Gyr for feature $\#12,13$ respectively.

Feature $\#14$ is unique because of its relative location close to younger terrain. It is located near the SPT margins in the leading hemisphere (around $300^\circ~\rm E$). However, there is a portion of more heavily cratered terrain surrounding this feature on the northern side (Figure \ref{fig:Crater_examples}). We take the older terrain age of $\sim 4~\rm Gyr$ as reasonable, despite the other nearby features ($\#11,13,15$) with younger estimated ages (ranging from $0.5~\rm Gyr$ to $3~\rm Gyr$). Due to the difference in crater density on either side of this feature, the terrain age estimate is more uncertain.

\subsection{Yield strength envelope method}
\label{supplement YSE}
We assume a conductive ice shell to set a thermal profile for the elastic layer. Assuming a conductive thermal profile for Enceladus' ice shell is reasonable given that there are large shell thickness variations \citep{hemingway_mittal_2019} which would be removed if the ice shell was convecting. Additionally, the thickness of Enceladus' ice shell limits the possibility of solid-state convection in the shell.

For a given heat flux, we can determine the temperature profile relative to the depth of the ice shell.
\be
\tag{A.1}
T(z) = T_s~\rm exp\left(\mathnormal{\frac{F c}{z}}\right)
\ee
Here, $T$ is temperature in K, $T_{s}$ is the surface temperature (set to be $70~\rm K$), $c$ is the temperature dependent thermal conductivity (equal to $651/T~\rm W/m$) \citep{petrenko_physics_1999}, $F$ is the surface heat flux in $\rm W/m^2$, and $z$ is the depth in the ice shell in m.

The YSE method assumes that the moment of the stress difference is equivalent between the elastic plate model and the brittle, elastic, ductile layer model. The moment is numerically integrated over the yield stress difference profile. To get the yield stress profile, we assume the strain rate of ice ($10^{-13}~\rm s^{-1}$), grain size (1 mm), friction coefficient (0.6), and basal-slip accommodated grain boundary sliding rheology ($n$=1.8) based on parameters experimentally determined by \cite{goldsby_ice_2001}.

For the elastic plate model, the moment is dependent on the elastic thickness, the curvature, Young's modulus, and Poisson's ratio \citep{mcnutt_1984,watts_isostasy_2001}.
\be
\tag{A.2}
M = \frac{K E\ {T_e}^3}{12\ (1 - \nu^2)}
\ee
Here, $M$ is the moment in $\rm N$, $E$ is Young's modulus of ice (taken to be $9~\rm GPa$), $\nu$ is Poisson's ratio (0.25), $K$ is the curvature of deformation in $\rm m^{-1}$, and $T_e$ is the elastic thickness in m. Equating the two moments produces a relationship between the elastic thickness, curvature, and surface heat flux (Figure \ref{fig:YSE}).

\section{Supplementary Files}
\subsection{Flexural Models}

\begin{figure}
    \centering
    \includegraphics[width=1.0\textwidth]{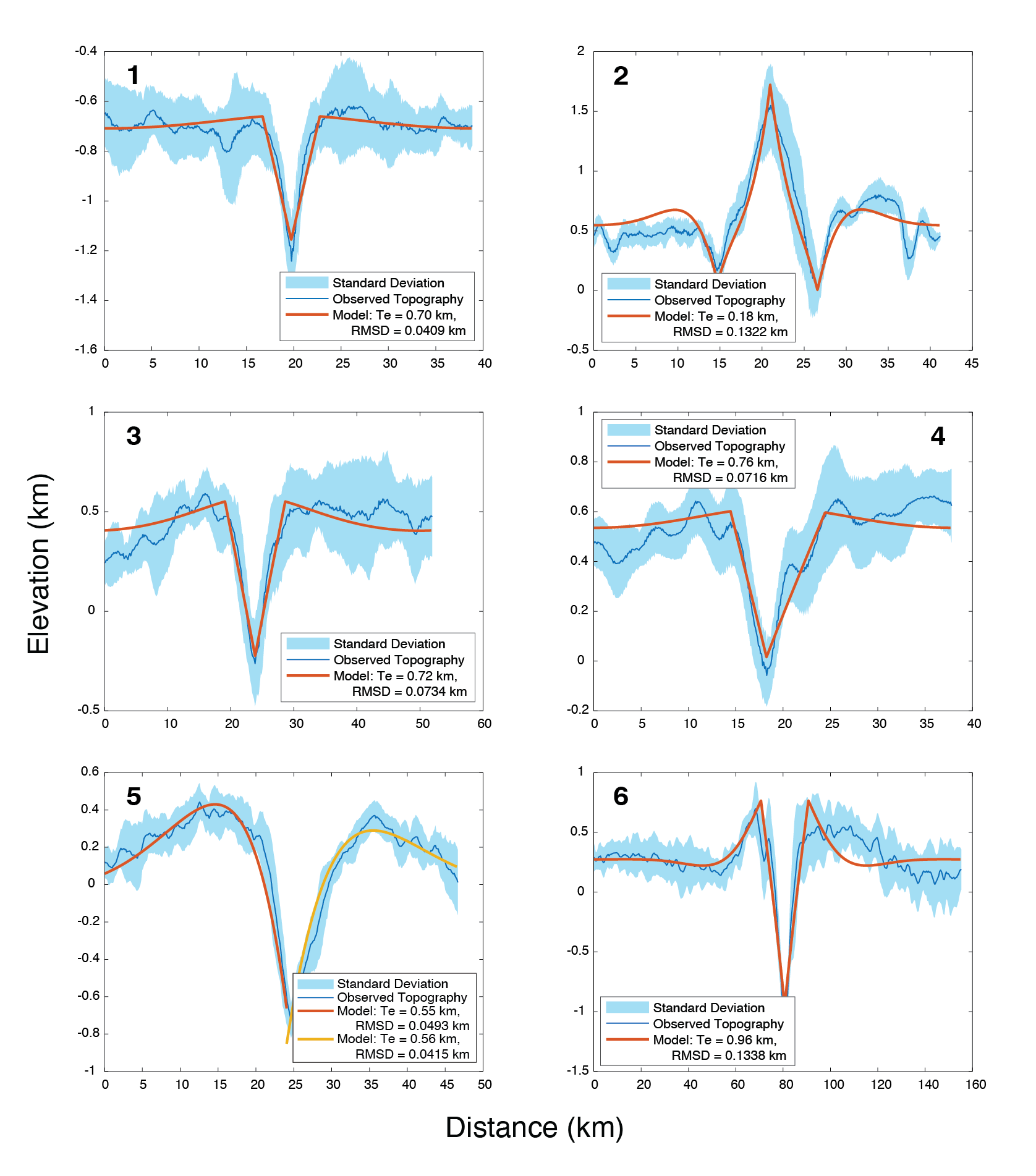}
    \caption{Best fit flexural model (orange) plotted with the averaged topography (blue) for features $\#1-6$ (Figure \ref{fig:FeatureMap}, Table \ref{tab:feature tab}). The blue shaded region represents the standard deviation of the topographic profiles.}
    \label{fig:Flexural_models_1}
\end{figure}

\begin{figure}
    \centering
    \includegraphics[width=0.9\textwidth]{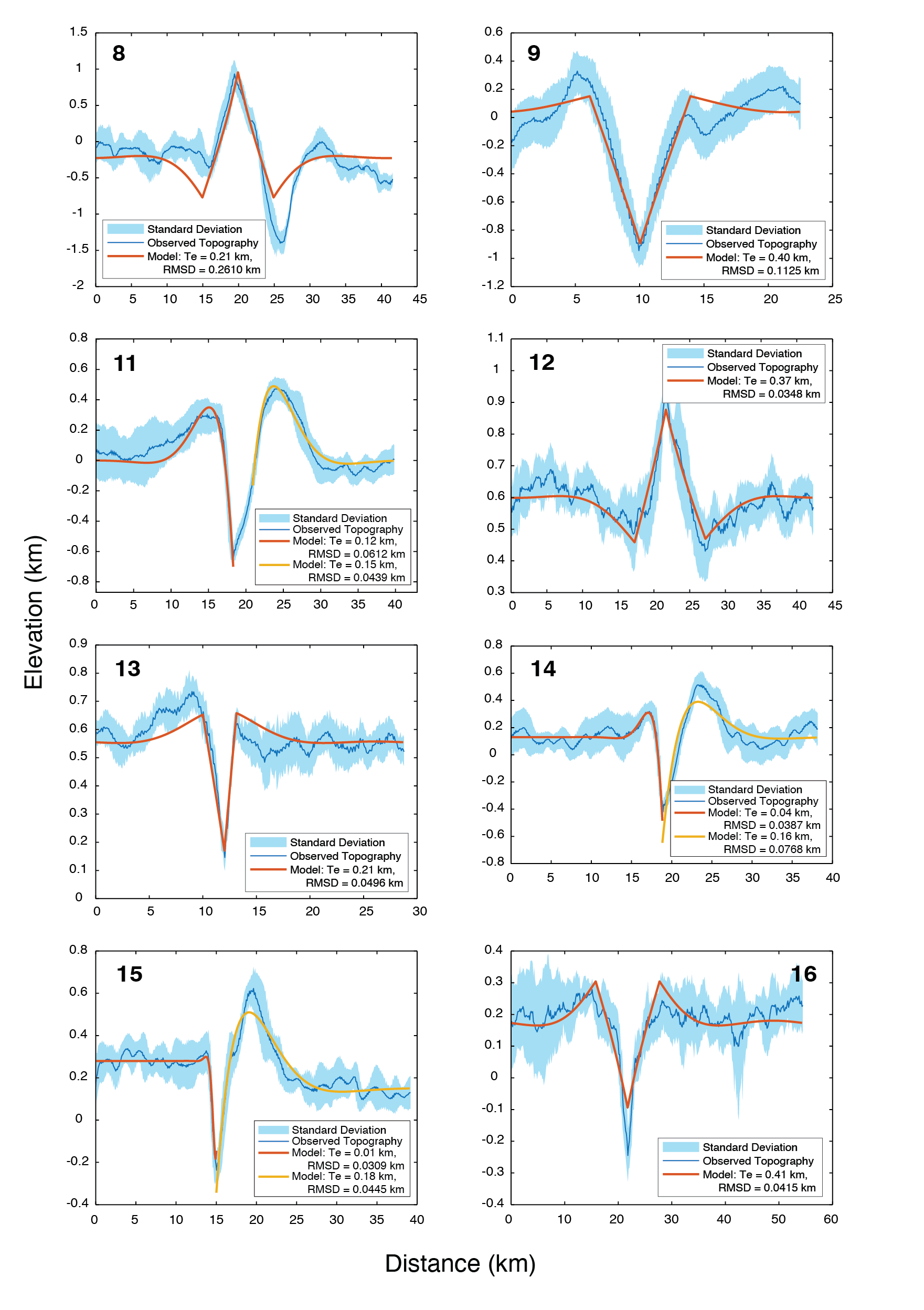}
    \caption{Best fit flexural model (orange) plotted with the averaged topography (blue) for features $\#8-9,11-16$ (Figure \ref{fig:FeatureMap}, Table \ref{tab:feature tab}). The blue shaded region represents the standard deviation of the topographic profiles.}
    \label{fig:Flexural_models_2}
\end{figure}

\end{document}